# Electron Mobility in Polarization-doped Al$_{0-0.2}$GaN with a Low Concentration Near 10$^{17}$ cm$^{-3}$

Mingda Zhu,[1,2,a)] Meng Qi,[2] Kazuki Nomoto,[1,2] Zongyang Hu,[1] Bo Song,[1] Ming Pan,[3] Xiang Gao,[3] Debdeep Jena,[1,2,4,5] Huili Grace Xing [1,2,4,5,b)]

[1]*School of Electrical and Computer Engineering, Cornell University, Ithaca, New York, 14853, USA*

[2]*Department of Electrical Engineering, University of Notre Dame, Notre Dame, Indiana, 46556, USA*

[3]*Department of Materials Science and Technology, Cornell University, Ithaca, NY 14853, USA*

[4]*Kavli Institute at Cornell for Nanoscale Science, Cornell University, Ithaca, NY 14853, USA*

[5]*IQE LLC, 265 Davidson Avenue, Somerset, New Jersey, 08873, USA*

In this letter, carrier transport in graded Al$_x$Ga$_{1-x}$N with a polarization-induced n-type doping as low as ~ 10$^{17}$ cm$^{-3}$ is reported. The graded Al$_x$Ga$_{1-x}$N is grown by metal organic chemical vapor deposition on a sapphire substrate and a uniform n-type doping without any intentional doping is realized by linearly varying the Al composition from 0% to 20% over a thickness of 600 nm. A compensating center concentration of ~10$^{17}$ cm$^{-3}$ was also estimated. A peak mobility of 900 cm$^2$/V·s at room temperature is extracted at an Al composition of ~ 7%, which represents the highest mobility achieved in n-Al$_{0.07}$GaN with a carrier concentration ~10$^{17}$ cm$^{-3}$. Comparison between experimental data and theoretical models shows that, at this low doping concentration, both dislocation scattering and alloy scattering are significant in limiting electron mobility; and that a dislocation density of <10$^7$ cm$^{-2}$ is necessary to optimize mobility near 10$^{16}$ cm$^{-3}$. The findings in this study provide insight in key elements for achieving high mobility at low doping levels in GaN, a critical parameter in design of novel power electronics taking advantage of polarization doping.

GaN has sparked intense research interest and found its way to a wide variety of electronic [1–3] and photonic [4] applications thanks to its unique properties, including the wide bandgap, polarization effects [5], and a high electron mobility. Although n-type doping in GaN is most commonly achieved by incorporating Si substitutional donors, it is also possible to introduce n-type doping in graded $Al_xGa_{1-x}N$ through polarization-induced internal electric field without involvement of intentional impurities (Pi-doping) [6]. The mobility of electrons generated by polarization-induced doping in AlGaN has been previously studied [7] at an electron concentration near $1.5\times10^{18}$ cm$^{-3}$, a carrier concentration relevant to RF devices [8], showing that alloy and phonon scattering are the main scattering mechanisms while neglecting dislocation scattering.

In this letter, the electron mobility in the Pi-doped AlGaN with a much lower electron concentration of ~ $10^{17}$ cm$^{-3}$ is studied, with an aim to implement polarization-induced doping in power electronic devices, where a low carrier concentration is desired: about $10^{15}$-$10^{16}$ cm$^{-3}$ for unipolar drift regions as in metal-oxide-semiconductor field effect transistors (MOSFETs) [9] or about $10^{17}$ cm$^{-3}$ for bipolar drift regions such as super-junctions [10]. The electron mobility is extracted experimentally using the split *C-V* method by combining current-voltage (*I-V*) and capacitance-voltage (*C-V*) characterizations of a field effect transistor (FET). Theoretical calculations of the electron mobility, taking into account ionized/neutral impurity, phonon, charged dislocation and alloy scattering, are carried out and compared with the experimental data. The comparison reveals that *for electron concentrations <$10^{17}$ cm$^{-3}$ in both Pi-doped AlGaN and Si-doped GaN, the key to achieving a high electron mobility is to reduce dislocation density to <$10^7$ cm$^{-2}$*. Pi-doped AlGaN is shown to have an advantage in electron mobility over Si-doped GaN at high electron concentrations due to the absence of impurity scattering.

The wafer in this study was grown by metalorganic chemical vapor deposition (MOCVD). The epitaxial growth started on a sapphire substrate with buffer layers, followed by an unintentionally doped (UID) semi-insulating Ga-face GaN with a thickness of 5 μm and a Pi-doped AlGaN with a thickness of $t \cong 600$ nm. Over this thickness, the Al composition was linearly increased from 0% to 20% toward the wafer surface. During the entire MOCVD growth, the Si precursor was not flown. The one-dimensional (1D) Poisson calculation incorporating polarization effects shows that a uniform electron concentration of $\rho_\pi \cong \frac{\rho_{surf} - \rho_{bulk}}{t} \cong 2 \times 10^{17}$ cm$^{-3}$ is expected due to the spontaneous and piezoelectric polarization in Al$_x$Ga$_{1-x}$N.

The secondary ion mass spectrometry (SIMS) scan of the sample shown in Fig. 1(a) confirms that the Al composition varies linearly from 20% at the surface to 0% at 600 nm depth. Both Si and H levels are at their detection limits of the SIMS measurement, and an increase of C and O (unintentional impurities) with increasing Al composition is observed. The morphology obtained from atomic force microscope (AFM) shown in Fig. 1(b) shows a very smooth surface after growth with a roughness root mean square (RMS) value smaller than 1 nm for the 5×5 μm$^2$ scan. Clear atomic steps are also observed in the AFM image, indicating good crystal quality of the Pi-doped AlGaN, although pit-like features are also visible, which are often observed on AlGaN surfaces [11].

Electron transport in Pi-doped AlGaN is characterized with Hall effect measurements at both room temperature (RT) and 77 K. The measured electron sheet concentration drops from $1 \times 10^{13}$ cm$^{-2}$ at RT to $5.9 \times 10^{12}$ cm$^{-2}$ at 77 K, while the electron mobility increases from 590 cm$^2$/Vs to 1540 cm$^2$/Vs at 77 K. Taking into account the surface depletion depth induced by an assumed surface barrier height of 1 eV, the electron bulk concentration is calculated to be $1.91 \times 10^{17}$ cm$^{-3}$ at RT and $1.17 \times 10^{17}$ cm$^{-3}$ at 77 K. Since carriers due to polarization-induced doping are activated by electric field, they are expected to exhibit a temperature independent behavior, as opposed to electrons thermally ionized from shallow donors like O that can be largely frozen out at 77 K, thus allowing an accurate measurement of the net polarization-induced doping concentration. The lower than expected carrier concentration of $1.17\times10^{17}$ cm$^{-3}$ at 77 K is attributed to the presence of compensating centers (e.g. C and Ga-vacancies) [12,13]. In Fig. 1(c), the temperature-dependent electron concentration of the Pi-doped AlGaN is shown, assuming an unintentional donor concentration of $N_D=6\times10^{16}$ cm$^{-3}$ with an activation energy of 34 meV [14] along with an unintentional deep acceptor concentration of $N_A=1\times10^{17}$ cm$^{-3}$. For comparison, the electron concentration of GaN with a Si doping (GaN:Si) concentration of $2\times10^{17}$ cm$^{-3}$ and the same unintentional impurities is also modeled and shown. The plot shows a close match between the model and experimental data. It again confirms that Pi-doping is much more resistant to freeze-out despite the presence of unintentional impurity dopants and compensating centers.

Metal-semiconductor FETs (MESFETs) were fabricated on the Pi-doped AlGaN sample to further characterize electron transport properties. The process flow includes source/drain ohmic contacts by regrowth using molecular beam epitaxy (MBE) [15] and Ti/Au metallization by e-beam evaporation, mesa device isolation with Cl$_2$-based inductive-coupled-plasma dry etching, followed by gate metal deposition of Ni/Au. The measured DC characteristics of the MESFETs are plotted in Fig. 2, while the device schematic is shown in the inset. The transfer curve of the MESFET (Fig. 2 (a)) shows an on/off ratio larger than $10^5$ at $V_{DS}=0.5$ V and the gate leakage current being low < 10 nA/mm throughout the measurement, benefitting from regrown ohmics [16]. The measured family curves are plotted in Fig. 2(b), showing an on-current larger than 0.2 A/mm.

Capacitance-Voltage (*C-V*) measurements were carried out at RT on the MESFETs with source/drain electrically grounded. The effective carrier concentration profile was then extracted by calculating the derivative of $1/C^2$, which is plotted in Fig. 3 along with the electron energy band diagram and the electron concentration from the 1D Poisson simulation, the *C-V* result and the depletion depth as a function of $V_{GS}$. The experimentally extracted doping profile shows a concentration around $10^{17}$ cm$^{-3}$ up to 450 nm deep in the sample followed by a decrease. At 600 nm below the surface, the effective doping concentration drops to ~ $10^{16}$ cm$^{-3}$, indicating that the depletion region in *C-V* measurement has reached the UID GaN layer underneath the Pi-doped AlGaN. No freeze-out effect is observed when the *C-V* measurement is conducted at -60 °C compared to RT. The effective doping concentration of ~$10^{17}$ cm$^{-3}$ is about 2 times smaller than the expected $2\times10^{17}$ cm$^{-3}$ from 1D Poisson calculations, which, as pointed out in the Hall effect measurement analysis, is most likely due to compensation effects from C incorporated in the epitaxy. A more detailed study of the compensation centers in the Pi-doped AlGaN will be carried out in future studies.

With the *C-V* measurement results and DC characteristics of the MESFET, electron low-field mobility could be extracted under the gradual channel approximation [7,17], i.e. the split *C-V* method. The circuit model used for mobility extraction is shown in Fig. 2. The contact resistance as well as the resistance from the access regions are accounted for using the results from transfer length method (TLM) measurements. The DC characteristics were carried out with a 20 μm long gate under a low drain to source voltage $V_{DS}$=0.1 V, corresponding to an effective electric field of ~50 V/cm, to minimize impact of non-uniform electric field along the FET channel, which is particularly important near the FET pinchoff. The drain current can be written as:

$$I_D(V_{GS}) = \int_{dep(V_{GS})}^{600\,nm} n(x) \times q \times \mu(x) \times E_{channel} dx \tag{1}$$

where $V_{GS}$ is the applied voltage between gate and source, *dep* is the depletion depth which is extracted from *C-V* results, $E_{channel} = \frac{\partial V}{\partial y}$ is the lateral electric field in the channel calculated from the voltage drop across the channel region, $n(x)$ and $\mu(x)$ are the effective doping concentration and the electron mobility at the depth *x*. $n(x)$ can be extracted from *C-V* results as shown in Fig. 3 (c). Taking the derivative with respect to $V_{GS}$ on both sides of Eq. (1):

$$\frac{dI_D(V_{GS})}{dV_{GS}} = -n(dep(V_{GS})) \times q \times \mu(dep(V_{GS})) \times E_{channel} \times \frac{d[dep(V_{GS})]}{dV_{GS}} \tag{2}$$

where all quantities except the electron mobility are either directly measured, or calculated from *I-V* or *C-V* measurement data. With the temperature-dependent DC characteristics and *C-V* results, the electron field effect mobility profile can be extracted, which is shown in Fig. 4 (a). The electron mobility increases with increasing depth, which is expected since alloy scattering reduces in the lower Al composition layers [7]. Beyond 450 nm depth, the electron mobility starts to decrease due to the decreasing electron concentration thus reduced screening of dislocation scattering, as will be explained next. Another trend shown in Fig. 4 (a) is the increase of mobility as the temperature decreases, which is a result of the decrease in phonon scattering [18]. The weighted average mobility is calculated by:

$$\mu_{average} = \frac{\int_{dep(0)}^{600\,nm} n(x) \times \mu(x) dx}{\int_{dep(0)}^{600\,nm} n(x) dx} \tag{3}$$

The maximum and average electron mobility values are shown in Fig. 4 (b). At RT the extracted electron field mobility ranges from 500 cm$^2$/Vs to 901 cm$^2$/Vs, with an average mobility of 724 cm$^2$/Vs. In comparison, the best electron Hall mobility reported in Si doped GaN grown on sapphire substrates with a similar electron concentration is 830 cm$^2$/Vs at RT [19]. This suggests that the average electron field mobility in Pi-doped AlGaN is comparable to that in Si doped GaN with comparable defect densities near a doping concentration of $10^{17}$ cm$^{-3}$. The monotonic increase in both average and maximum electron mobility as temperature decreases is observed in Fig. 4(b), with the maximum mobility at 213 K exceeding 2000 cm$^2$/V. The extracted mobility of 900 cm$^2$/Vs at n~$10^{17}$ cm$^{-3}$ in Pi-doped Al$_{0.07}$GaN is among the highest mobility obtained in n-Al$_{0.07}$GaN at a similar doping level.

To understand the limiting factors of electron mobility in Pi-doped AlGaN, a RT electron mobility model is constructed [20], taking into consideration acoustic and optical phonon scattering, alloy scattering, ionized/neutral impurity scattering as well as dislocation scattering [21,22]. For improved accuracy, the Al composition and impurity densities of carbon and oxygen from the SIMS measurement, the electron concentration from the *C-V* measurements are employed in the mobility modeling shown in Fig. 5(a). A dislocation density of $10^9$ cm$^{-2}$ is used, which is typical in GaN on sapphire; since not all dislocations are charged and charged dislocations induce a higher scattering rate, a charge occupation probability also needs to be determined. A good fit using a charge occupation probability of 55% was found between the measured and calculated mobility, as shown in Fig. 5(a); therefore, this value is used throughout all the modeling in this work and the unoccupied dislocations are treated as neutral impurities. The electron component-mobilities limited by various scattering mechanisms for Al$_{0.07}$GaN (peak mobility observed in this work) with a dislocation density of $10^9$ cm$^{-2}$ are modeled and plotted in Fig. 5(b) along with the total electron mobility calculated using the Matthiessen's rule [20]; impurity scattering off carbon and oxygen is neglected in Fig. 5(b) in order to delineate the effect of dislocation scattering. Experimental data from this work and Ref. [7] are also plotted for comparison. It is seen that dislocation scattering is the only scattering mechanism that has a significant electron concentration dependence; dislocation-scattering-limited electron mobility decreases with decreasing electron concentration due to weakening of the screening effect at lower electron concentrations. The effect of dislocation scattering can be understood as follows: charged dislocations scatter carriers largely like ionized impurities; a scatter-center volume concentration induced by charged dislocations can be estimated by $N_{dis}/c \sim 10^9$ cm$^{-2}$/0.5 nm = $2\times10^{16}$ cm$^{-3}$, where $N_{dis}$ is the dislocation density and $c$ is the unit cell height of GaN along the [0001] direction. For $n>>N_{dis}/c$, the effect of dislocations is sufficiently screened, as in the Al$_{0.15->0.07}$GaN layer in this work (~$10^{17}$ cm$^{-3}$) and Ref.7 (~$10^{18}$ cm$^{-3}$); for $n \leq N_D/c$, dislocation scattering is not negligible, thus leading to lower mobility in the Al$_{0.07->0}$GaN layer. If dislocation scattering were neglected, on the other hand, the calculated mobility at low carrier concentrations would be much higher, as is the case presented in Ref.7. It is also believed that in Ref.7 a large error is associated with the experimental extraction of the mobility in the Al$_{0.07->0}$GaN layer since it was extracted near the FET pinchoff but a constant current thus a high applied field was used in calculating the carrier mobility.

To further illustrate the impact of dislocations in Pi-doped and Si-doped n-GaN, the theoretical electron mobility for Pi-doped AlGaN is plotted in Fig 6 (a) as a function of electron concentration with two dislocation densities and at two Al compositions. The modeled electron mobility in Pi-doped Al$_{0.05}$GaN is then compared with Si-doped GaN, shown in Fig. 6 (b). In these models, all the aforementioned scattering mechanisms are considered; however, impurity scattering is excluded in Pi-doped AlGaN assuming an ideal epitaxy, while in Si doped GaN, alloy scattering is excluded and impurity scattering is included using an activation energy of Si: $\Delta E_D \sim 20$ meV. It can be found in Fig. 6 (a) that at a low dislocation density of $10^7$ cm$^{-2}$, the electron mobility in Pi-doped GaN is largely independent of electron concentration. This is because of the dominance of alloy scattering, which is independent of electron concentration in nondegenerate regime [20]. An increase in dislocation density from $10^7$ cm$^{-2}$ to $10^9$ cm$^{-2}$, results in an increasingly high dependence of electron mobility on the electron concentration. The effect of dislocation scattering is more pronounced in Al$_{0.07}$GaN compared to Al$_{0.15}$GaN because of a lower alloy scattering in Al$_{0.07}$GaN. The significant effect of dislocation scattering is also observed in Si-doped GaN with low electron concentrations, also compared to Pi-doped GaN in Fig. 6(b). At low carrier concentrations (<$2\times10^{17}$ cm$^{-3}$), the presence of dislocations severely degrades mobility in layers doped in both schemes. The experimentally reported mobility values (symbols) for $N_{dis}$ ~$10^9$ cm$^{-2}$ are also included. The difference between the experimental values and the modeled values can be attributed to experimental errors as well as scattering due to compensating point defects, which is not included in the models.

For $n > 2\times10^{17}$ cm$^{-3}$, the electron mobility is severely degraded by impurity scattering in GaN:Si [23,24]; on the contrary, the electron mobility in Pi-doped AlGaN remains high for high electron concentrations since there is no impurity scattering. At low electron concentrations, dislocation scattering has a significant impact on electron mobility in both Pi-doped AlGaN and Si-doped GaN. Thus, the key to improving electron mobility for electron concentration < $10^{17}$ cm$^{-3}$ is to reduce $N_{dis}$. As a much lower $N_{dis}$ (< $10^7$ cm$^{-2}$) is achievable in today's bulk GaN substrates and the subsequent epitaxial layers [25,26] than GaN on SiC/sapphire substrates, it is feasible to achieve improved electron mobility in Pi-doped AlGaN.

In summary, linearly graded Al$_x$Ga$_{1-x}$N is grown by MOCVD on sapphire substrates with a polarization induced doping of a low and uniform electron concentration near $10^{17}$ cm$^{-3}$. A peak electron mobility of ~900 cm$^2$/Vs at RT is extracted in Al$_{0.07}$GaN. By comparing the experimental data to carrier transport models, dislocation scattering is found to be the significant factor limiting electron mobility when the electron concentration is below $10^{17}$ cm$^{-3}$. At low electron concentrations <$10^{17}$ cm$^{-3}$, much improved electron mobility can be expected in epitaxial layers grown on bulk GaN substrates with a dislocation density lower than $10^7$ cm$^{-2}$.

This work was partly supported by the ARPAe SWITCHES project monitored by Drs. Tim Heidel and Isik C. Kizilyalli.

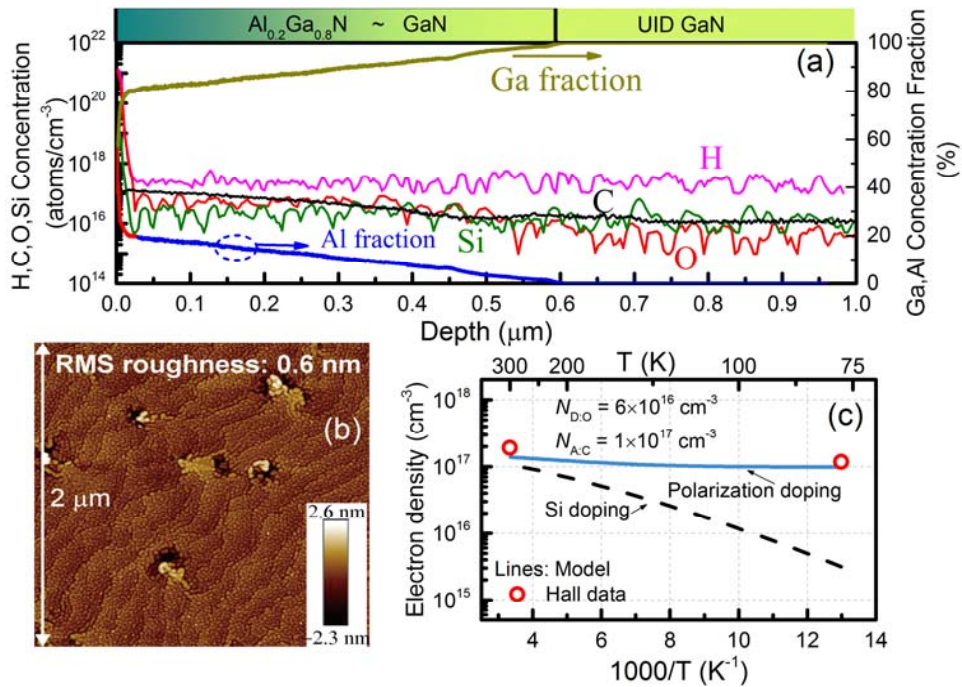

FIG. 1. (a) Impurity concentration (H, C, O Si) and Ga/Al composition profiles of the MOCVD grown Pi-doped AlGaN from secondary ion mass spectrometry (SIMS) measurement. (b) AFM images of 5×5 μm² scan on the Pi-doped AlGaN. Clear atomic steps and smooth surface is observed. (c) Electron concentration obtained from Hall effect measurements at RT and 77 K along with modeled electron concentration versus temperature for Pi-doped and Si-doped GaN, confirming electrons in the AlGaN are indeed induced by polarization doping.

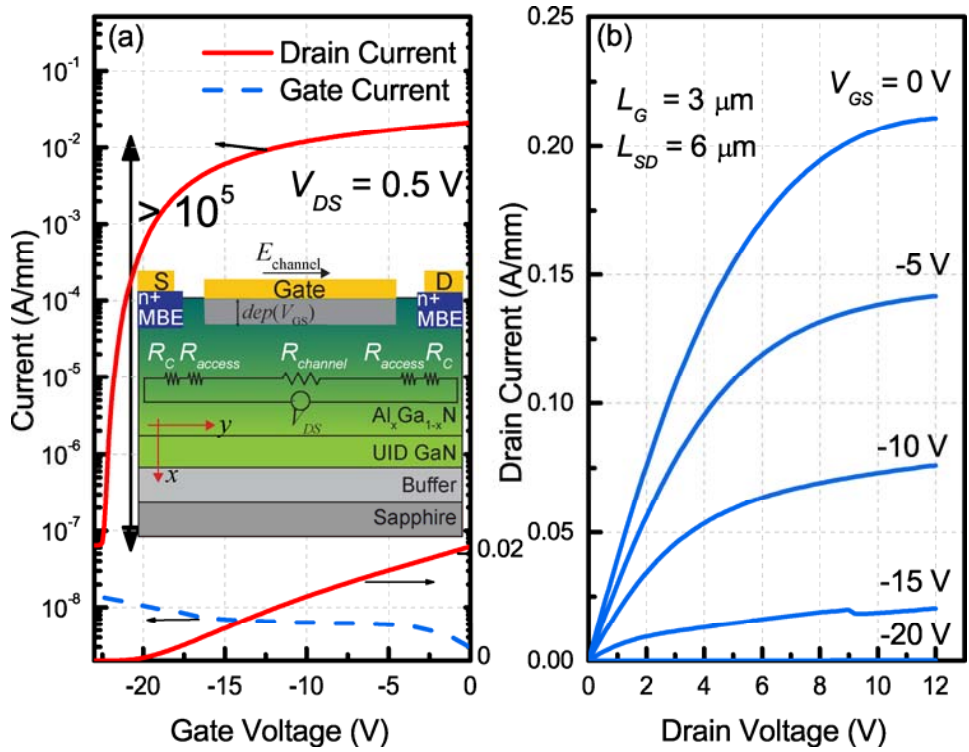

FIG. 2 . DC characteristics of the MESFETs with Pi-doped channel: (a) transfer curves and (b) family $I_D$-$V_{DS}$ curves. The inset in (a) shows schematic of the fabricated MESFET. The MESFETs are used to extract electron field mobility in the Pi-doped AlGaN channel.

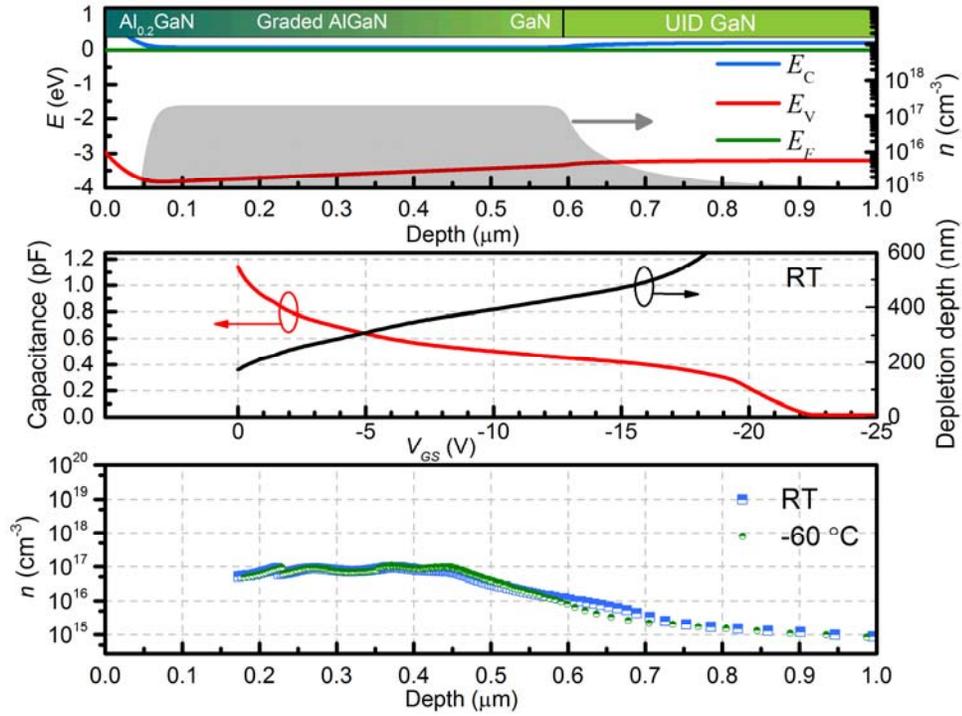

FIG. 3. (a) Energy band diagram and electron concentration profile along the epitaxial growth direction from 1-D Poisson simulations. (b) Capacitance/depletion depth as a function of $V_{GS}$ from C-V measurements of the MESFET. (c) Extracted doping concentration profile at RT and -60 °C.

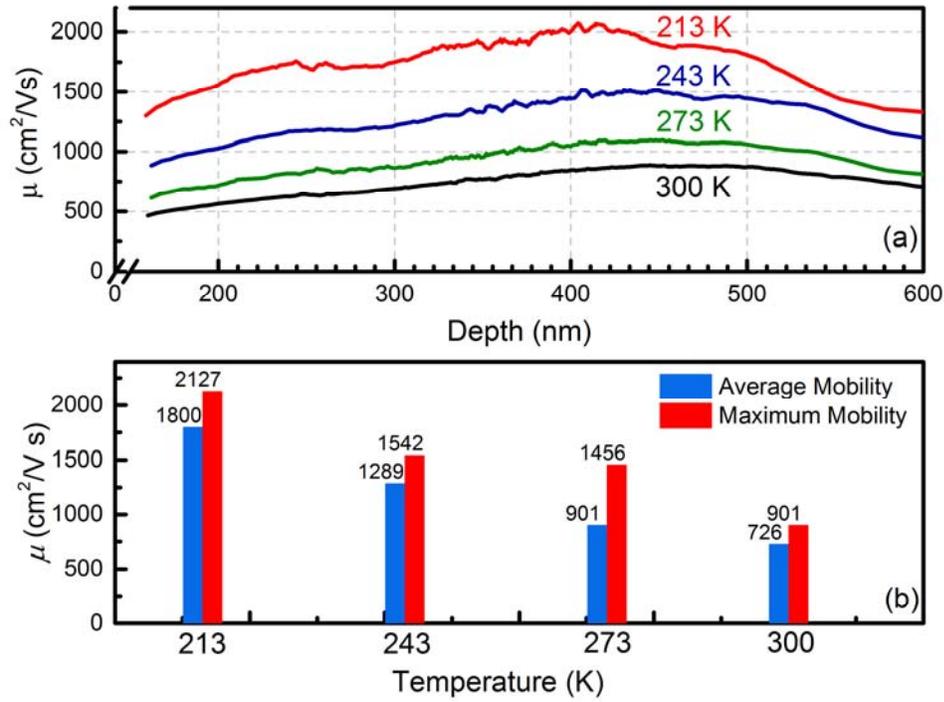

FIG. 4. (a) Extracted electron mobility as a function of depth from the sample surface at various temperatures. (b) Average/maximum electron mobility extracted in the Pi-doped AlGaN versus temperature.

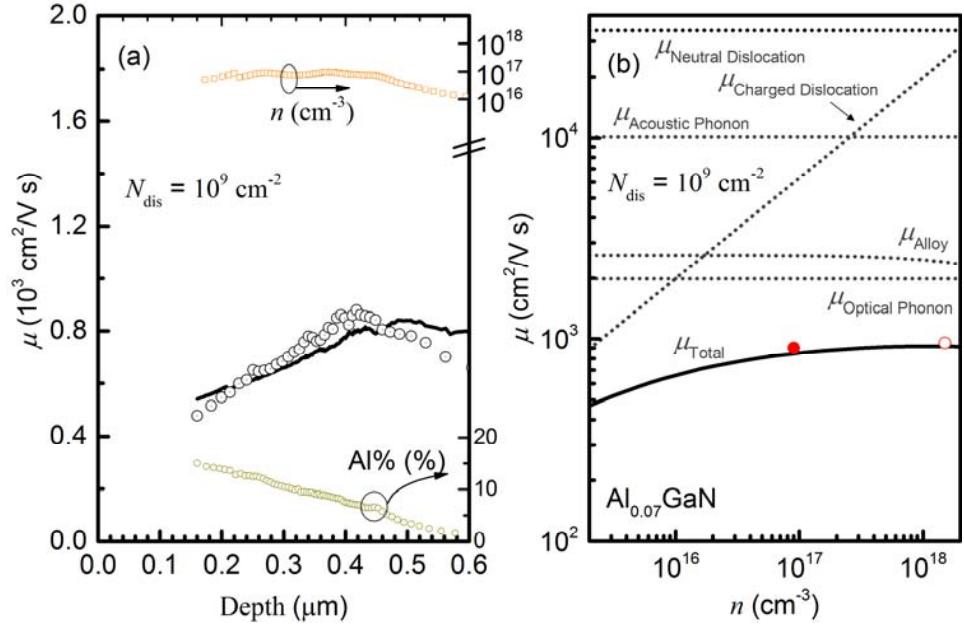

FIG. 5. (a) Experimentally extracted and calculated electron mobility for the Pi-doped AlGaN as a function of depth from the sample surface along with the measured Al composition and electron concentration. The calculations take into account optical phonon scattering, alloy scattering, dislocation scattering, impurity scattering with the Al, Si and C concentrations all determined from SIMS. (b) Modeled electron mobilities vs. electron concentration in Pi-doped $Al_{0.07}GaN$ with a dislocation density of $10^9$ cm$^{-2}$. The $Al_{0.07}GaN$ is assumed to be free of impurities otherwise. Experimental data from this work (solid red circle) and from Ref. 7 (open red circle). The occupation probability of dislocation is assumed to be 55% in all modeling results and the unoccupied dislocations are treated as neutral impurities.

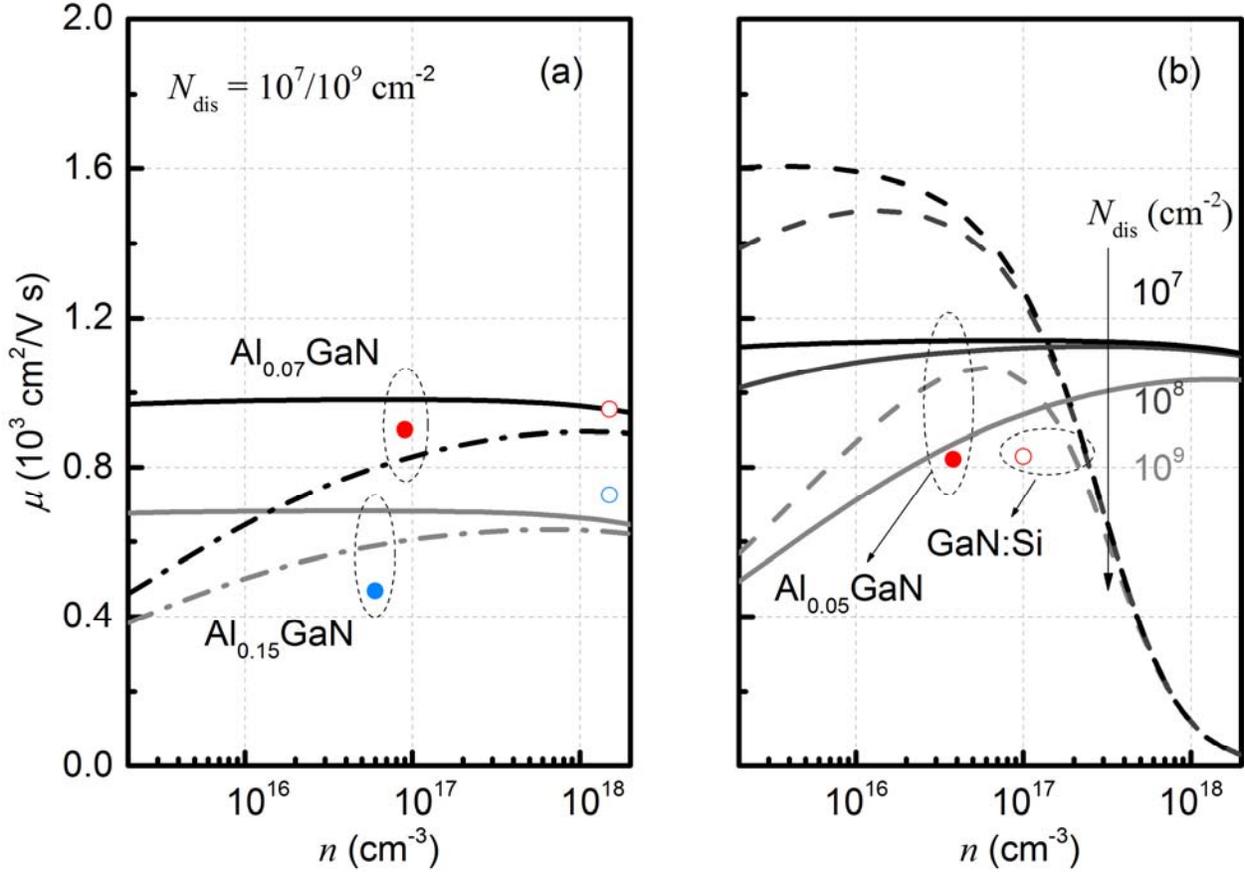

FIG. 6 Impact of dislocations on electron mobility. (a) Modeled electron mobility in Pi-doped Al$_{0.07}$GaN and Al$_{0.15}$GaN assuming an $N_{dis}$ of $10^7$ cm$^{-2}$ (solid lines) and $10^9$ cm$^{-2}$ (dash-dot lines) with the otherwise same assumptions in Fig. 5(b). Also shown are the experimentally measured electron mobility in AlGaN from this work (solid circles) and Ref. [7] (open circles). (b) Modeled electron mobility in Pi-doped Al$_{0.05}$GaN and Si-doped GaN as a function of the electron concentration at a few dislocation densities. Also shown are the experimentally measured electron mobility in Al$_{0.05}$GaN (solid circle, this work) and that in GaN:Si (open circle, Ref.[23]) in GaN on sapphire with similar $N_{dis}$ values. Not shown in the plot, the electron mobility in GaN grown on bulk GaN with $N_{dis}$ <$10^7$ cm$^{-2}$ has been reported to be ~1500 cm$^2$/Vs near a concentration ~$10^{16}$ cm$^{-2}$ [26] The finite differences between the models and experiments is largely attributed to experimental errors as well as the existence of other point defects and defect complexes that are not considered in the models.